\documentclass[aps,prl,amssymb,reprint,twocolumn,superscriptaddress,longbibliography]{revtex4-1}

\usepackage[ansinew]{inputenc}
\usepackage{graphicx}
\usepackage[breaklinks=false,colorlinks=true,linkcolor=blue,urlcolor=blue,citecolor=blue]{hyperref}
\usepackage{setspace}
\usepackage[ansinew]{inputenc}
\usepackage{caption}
\usepackage[labelfont=bf]{caption}
\usepackage{graphicx}
\usepackage{amsmath}
\usepackage{color}
\usepackage{lineno}
\usepackage{ulem}
\begin{document}

\title{Compact intense extreme-ultraviolet source} 

\author{B. Major}
\affiliation{ELI-ALPS, ELI-HU Non-Profit Ltd., Wolfgang Sandner utca 3., Szeged 6728, Hungary}
\affiliation{Department of Optics and Quantum Electronics, University of Szeged, D\'om t\'er 9, Szeged 6720, Hungary}
\author{O. Ghafur}
\affiliation{Max-Born-Institut, Max-Born-Strasse 2A, 12489 Berlin, Germany}
\author{K. Kov\'acs}
\affiliation{National Institute for Research and Development of Isotopic and Molecular Technologies, Donat str. 67-103, 400293 Cluj-Napoca, Romania}
\author{K. Varj\'u}
\affiliation{ELI-ALPS, ELI-HU Non-Profit Ltd., Wolfgang Sandner utca 3., Szeged 6728, Hungary}
\affiliation{Department of Optics and Quantum Electronics, University of Szeged, D\'om t\'er 9, Szeged 6720, Hungary}
\author{V. Tosa}
\affiliation{National Institute for Research and Development of Isotopic and Molecular Technologies, Donat str. 67-103, 400293 Cluj-Napoca, Romania}
\author{M. J. J. Vrakking}
\affiliation{Max-Born-Institut, Max-Born-Strasse 2A, 12489 Berlin, Germany}
\author{B. Sch\"utte}
\affiliation{Max-Born-Institut, Max-Born-Strasse 2A, 12489 Berlin, Germany}
\email{schuette@mbi-berlin.de}
\date{\today}

\begin{abstract}
High-intensity laser pulses covering the ultraviolet to terahertz spectral regions are nowadays routinely generated in a large number of laboratories. In contrast, intense extreme-ultraviolet (XUV) pulses have only been demonstrated using a small number of sources including free-electron laser facilities~\cite{ackermann07, shintake08, allaria12} and long high-harmonic generation (HHG) beamlines~\cite{takahashi02,ravasio09,tzallas11,schutte14,manschwetus16,bergues18}. Here we demonstrate a concept for a compact intense XUV source based on HHG that is focused to an intensity of $2 \times 10^{14}$\,W/cm$^2$, with a potential increase up to $10^{17}$\,W/cm$^2$ in the future. Our approach uses tight focusing of the near-infrared (NIR) driving laser and minimizes the XUV virtual source size by generating harmonics several Rayleigh lengths away from the NIR focus. Accordingly, the XUV pulses can be refocused to a small beam waist radius of 600\,nm, enabling the absorption of up to four XUV photons by a single Ar atom in a setup that fits on a modest (2\,m) laser table. Our concept represents a straightforward approach for the generation of intense XUV pulses in many laboratories, providing novel opportunities for XUV strong-field and nonlinear optics experiments, for XUV-pump XUV-probe spectroscopy and for the coherent diffractive imaging of nanoscale structures.
\end{abstract}

\maketitle

Nonlinear optical techniques have widespread applications including frequency mixing~\cite{franken61, bass62, ferray88}, Raman amplification~\cite{claps03}, Kerr-lens modelocking~\cite{spence91} and self-phase modulation~\cite{alfano70, brabec00}. In the long-wavelength range, the development of compact and efficient secondary terahertz sources with microjoule energies~\cite{yeh07} has boosted the field of nonlinear terahertz optics~\cite{hoffmann11}. The situation is different in the XUV and X-ray wavelength range, where the generation of high pulse energies in combination with high intensities has only been demonstrated at a limited number of facilities including free-electron lasers~\cite{ackermann07, shintake08, allaria12} and at long ($\geq 10$\,m) HHG beamlines~\cite{takahashi02,ravasio09,tzallas11,schutte14,manschwetus16,bergues18}. While first nonlinear XUV optics experiments including multiphoton absorption~\cite{sorokin07,nayak18,senfftleben20}, four-wave mixing~\cite{bencivenga15} and XUV-XUV pump-probe spectroscopy~\cite{ding19} have been performed, the large sizes and complexity of existing experimental setups using intense XUV pulses impede faster progress in this field, which would benefit from more compact and less complex setups. Such sources would furthermore pave the way to the application of powerful techniques like attosecond-pump attosecond-probe spectroscopy~\cite{tzallas11,takahashi13} and coherent diffractive imaging of nanoscale structures and nanoparticles~\cite{bogan08,ravasio09,rupp17} in a much larger number of laboratories than is currently the case.

Our concept for a compact intense XUV source is based on HHG, which has traditionally been performed at or close to the focal plane of the fundamental laser~\cite{ferray88} because of the high driving light intensities ($\approx 10^{14}$\,W/cm$^2$) that are required for efficient HHG. After more powerful driving lasers became available, loose-focusing geometries were adopted~\cite{takahashi02,ravasio09,tzallas11,schutte14,manschwetus16,bergues18}, in which the fundamental laser is focused by a lens or a spherical mirror with a long focal length on the order of 10\,m. This increases the focus size of the fundamental laser, leading to a large volume from which high harmonics are emitted and resulting in XUV pulses with microjoule energies~\cite{takahashi02,hergott02, nayak18}. A crucial disadvantage of this approach is its complexity and its intrinsic requirement of a large laboratory. Moreover, this approach does not readily lead to higher XUV intensities, since the XUV source size grows proportionally to the focal length that is used~\cite{heyl16}, and source demagnification factors that can be achieved using focusing optics are finite.

\begin{figure}[htb]
 \centering
  \includegraphics[width=8.6cm]{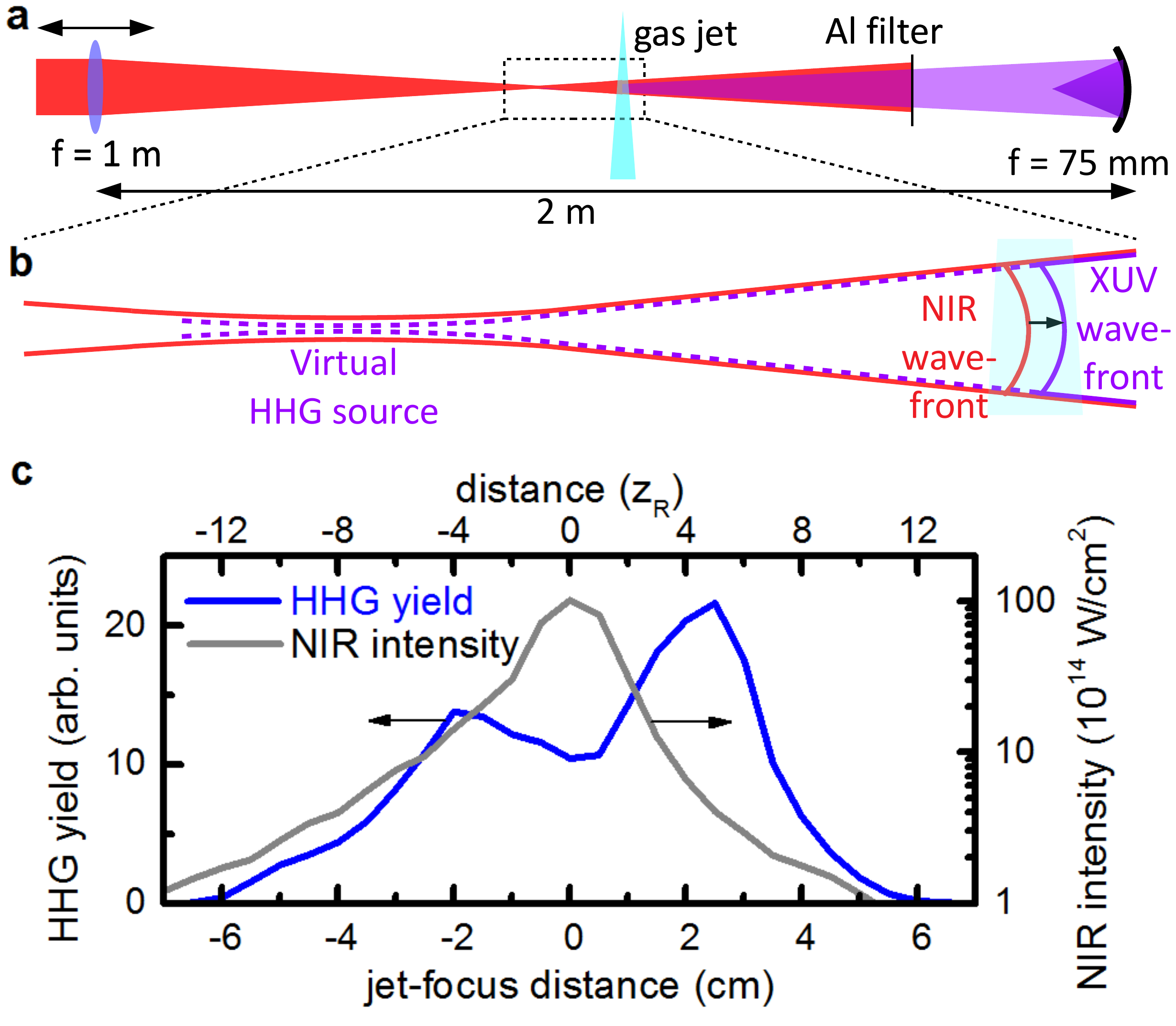}
 \caption{\label{figure_setup} \textbf{Compact intense XUV source.} \textbf{a,} Experimental setup: NIR driving pulses with a duration of 40\,fs and a pulse energy of up to 16\,mJ are focused using a spherical lens with $f=1$\,m. HHG takes place in a high-pressure gas jet that is placed in the converging or diverging NIR beam. An Al filter attenuates the NIR pulses after HHG, and the XUV pulses are focused to high intensities by a spherical mirror with $f=75$\,mm. \textbf{b}, Zoom into the generation region, visualizing the transfer of curved wavefronts from the fundamental to the harmonic beam. This leads to a virtual HHG source size which is significantly smaller than the NIR focus size. \textbf{c,} HHG yield using Kr at a backing pressure of 4~bar as a function of the distance between the jet and the NIR focus (blue curve). The maximum XUV photon yield is observed when the jet is placed 2.5\,cm behind the NIR focus. The relative NIR peak intensity as a function of the jet-focus distance (gray dotted curve) was determined using a camera, while the absolute NIR peak intensity was determined at the NIR focal plane using the measured NIR beam waist radius, pulse duration and pulse energy. At 2.5\,cm the NIR peak intensity is $4 \times 10^{14}$\,W/cm$^2$, i.e. 25 times smaller than at the focus.}
\end{figure}

In order to increase the focused XUV intensity in a compact setup, we propose to generate high harmonics several Rayleigh lengths away from the focal plane of the driving laser, employing a focusing element with a relatively short focal length of 1\,m (see Fig.~1a). As schematically shown in Fig.~1b, this leads to a situation where the driving laser wavefronts in the HHG medium are curved, resulting in curved wavefronts of the generated XUV pulses~\cite{wikmark19}. Due to the short wavelengths of the generated harmonics, this curved wavefront is accompanied by a virtual HHG source size that is much smaller than the focus size of the driving laser, as we will demonstrate later. Further demagnification of the HHG source size (here, using a spherical B$_4$C-coated mirror with $f=75$\,mm) results in a small XUV focus size and a high XUV intensity that can be used for experiments. In the experiments to be discussed in this paper, the entire XUV beamline has a length of 2\,m, which is comparable to or even smaller than most standard HHG beamlines. 

To demonstrate the applicability of this scheme, we performed an experiment at the Max-Born-Institut (MBI), where high harmonics were generated in a Kr gas jet (1.5\,mm length) that was operated at a backing pressure of 4\,bar, using 40~fs near-infrared (NIR) driving pulses with a central wavelength of 800\,nm and a pulse energy of 16\,mJ. Fig.~1c shows the HHG yield as a function of the distance between the NIR focal plane and the gas jet. As a general feature we observe two maxima of the HHG yield, one when the jet is behind the NIR focus and another one when the jet is in front of the NIR focus. For the specific parameters used in Fig.~1c, the curve has a maximum at $2.5$\,cm, meaning that the jet is placed 2.5\,cm --- or about 5 Rayleigh lengths --- behind the NIR focus. The NIR intensity at this position is $4\times 10^{14}\,$W/cm$^2$, which is much smaller than the NIR intensity at the focus ($1\times 10^{16}\,$W/cm$^2$). This corresponds to a regime where the NIR beam propagation and the propagation of the XUV beam resulting from HHG can be well approximated by geometrical optics. Using this scheme, an XUV pulse energy of 0.3\,$\mu$J was measured using an XUV photodiode (AXUV100G), which is comparable to the results obtained with two long HHG beamlines available at the MBI~\cite{schutte14,senfftleben20}.

\begin{figure}[tb]
 \centering
  \includegraphics[width=8.6cm]{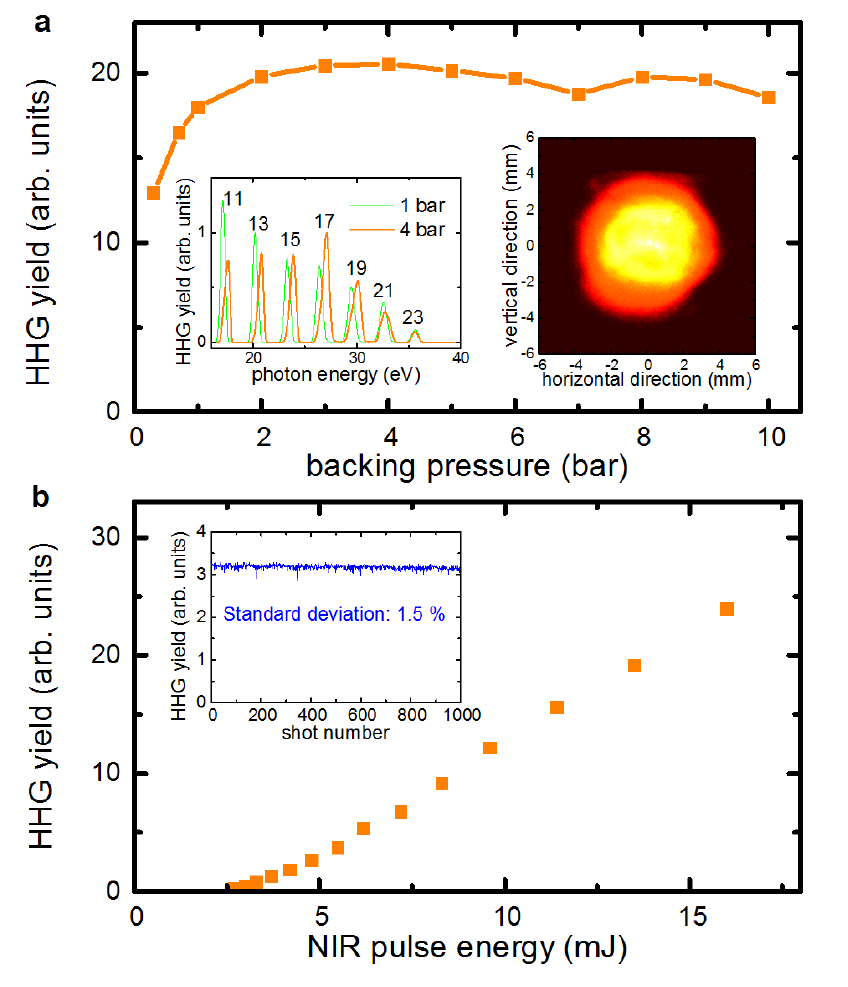}
 \caption{\label{figure_HHGproperties} \textbf{Measured XUV properties.} \textbf{a,} HHG yield in Kr as a function of the backing pressure at a fixed jet-focus distance of $2.5$\,cm. The left inset shows HHG spectra at backing pressures of 1\,bar and 4\,bar, exhibiting a clear blueshift in the latter case. The XUV beam profile measured 50\,cm behind the jet using a backing pressure of 4\,bar is shown in the inset on the right side. \textbf{b}, Using a backing pressure of 4\,bar, the HHG yield increases monotonically with the NIR pulse driving energy. The very good shot-to-shot stability of the HHG yield over 1000 single shots is shown in the inset.}
\end{figure}

We will show in the following that HHG far away from the NIR focus exhibits a number of favorable properties which are beneficial for applications. As depicted in Fig.~2a, variation of the backing pressure results in an HHG yield that is almost constant for backing pressures $>1$\,bar, making the optimization of HHG straightforward. Corresponding HHG spectra at backing pressures of 1\,bar and 4\,bar are shown in the left inset and demonstrate that the individual harmonics are blueshifted when the pressure is increased, which provides a possibility to tune the HHG spectra. This blueshift is a consequence of propagation effects of the NIR driving laser in the gas jet~\cite{major20}. The XUV beam profile measured 50\,cm from the gas jet is depicted in the inset on the right side of Fig.~2a. As a direct consequence of generating harmonics far away from the NIR focus, we find that the corresponding full-width half maximum divergence matches the NIR divergence of 10\,mrad. Fig.~2b shows that the HHG yield increases monotonically with increasing NIR pulse energy, demonstrating that the presented HHG scheme is scalable. A further advantage is that the shot-to-shot fluctuations are low, making this a very robust source. As shown in the inset of Fig.~2b, a standard deviation of 1.5\,$\%$ was measured over 1000 single shots.    

\begin{figure}[tb]
 \centering
  \includegraphics[width=8.6cm]{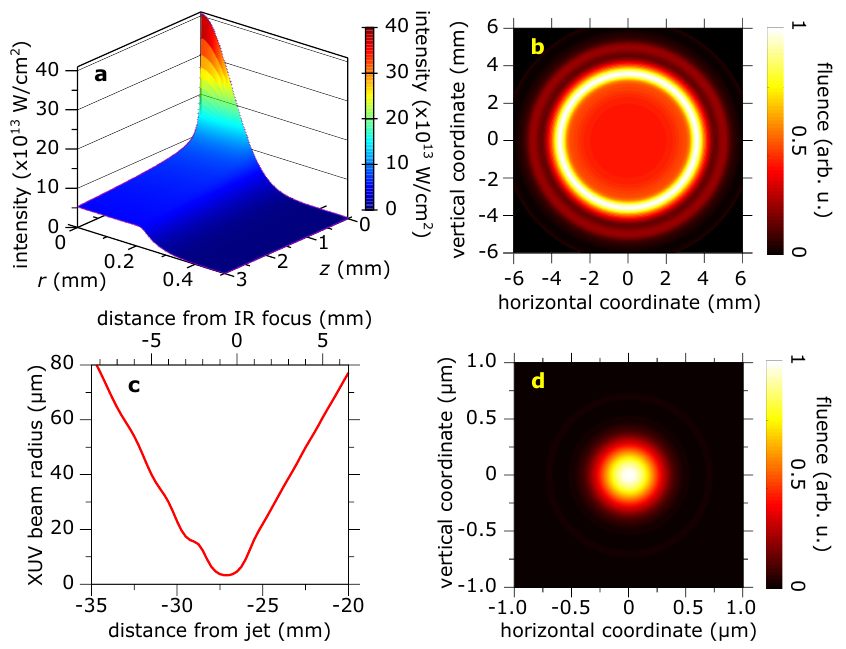}
 \caption{\label{figure_simulation} \textbf{Simulation of HHG far away from the NIR focus.} \textbf{a,} Evolution of the NIR beam profile during propagation in the Kr gas jet. \textbf{b,} Simulated XUV beam profile 50 cm away from the gas jet.
\textbf{c,} The beam radius of the backpropagated XUV beam as a function of distance from the gas jet shows that the position of the virtual XUV source almost coincides with the NIR focal plane, which is consistent with the predictions of the modeling performed in Ref.~\cite{wikmark19}. \textbf{d,} 
 Simulated XUV beam profile at the image plane of the virtual HHG source after focusing by a $75\,\mathrm{mm}$ focal length spherical mirror.  
 }
\end{figure}

To better understand HHG far away from the driving laser focus, we performed simulations (see Methods for details). Fig.~3a shows how the NIR pulse is reshaped in the gas medium: The initial Gaussian spatial profile of the driving laser is modified to an almost flat-top NIR profile after propagating through the jet as a result of absorption (including losses due to ionization~\cite{geissler99}). In this way, an NIR intensity of $\sim 6 \times 10^{13}$\,W/cm$^2$ is achieved over a large volume. As a consequence, a relatively high XUV flux is observed in the simulations, which is comparable to the XUV flux that we have obtained in a simulation using an upscaled HHG source, where an NIR focal length of 9\,m and an extended gas medium with a length of 20\,mm were used. The corresponding XUV beam profile at a distance of 50\,cm from the gas jet presented in Fig.~3b exhibits a divergence similar to the experiment. This beam profile has an annular structure, which can be less or more pronounced depending on the specific parameters used in the simulations. We note that an annular beam profile was also observed experimentally in certain conditions, but for the current experiments we chose to optimize the beam profile to be more homogeneous (see inset of Fig.~2a). Difference in the XUV beam profiles might also be explained by the fact that an ideal NIR Gaussian beam was used in the simulation, whereas the NIR beam used in the experiment was not an ideal Gaussian beam. As depicted in Fig.~3c, backpropagation of the simulated XUV pulses --- centered at 55\,nm wavelength --- shows that the virtual HHG source is located close to the NIR focal plane and has a beam waist radius of only 3.5\,$\mu$m (Fig.~3c), which is significantly smaller than the simulated NIR beam waist radius of 30\,$\mu$m. Demagnification of the virtual XUV source using a spherical XUV focusing mirror with $f=75$\,mm placed 70\,cm behind the NIR focus results in an XUV beam waist radius of only $350$\,nm, which is substantially smaller compared to the values achieved in loose-focusing HHG setups~\cite{ravasio09, schutte14, manschwetus16, rupp17, nayak18}. Using the calculated beam waist radius in combination with the calculated XUV pulse duration of $\Delta t=25$\,fs and using a pulse energy of $E=30$\,nJ as available in the experiment (taking into account an aluminium (Al) filter transmission of 40\,$\%$ and a focusing mirror reflectivity of 25\,$\%$), the theoretically achievable XUV peak intensity is estimated as $I_{peak} = 2E/(\Delta t \pi w_0^2)=8\times10^{14}$\,W/cm$^2$. 

\begin{figure}[tb]
 \centering
  \includegraphics[width=8.6cm]{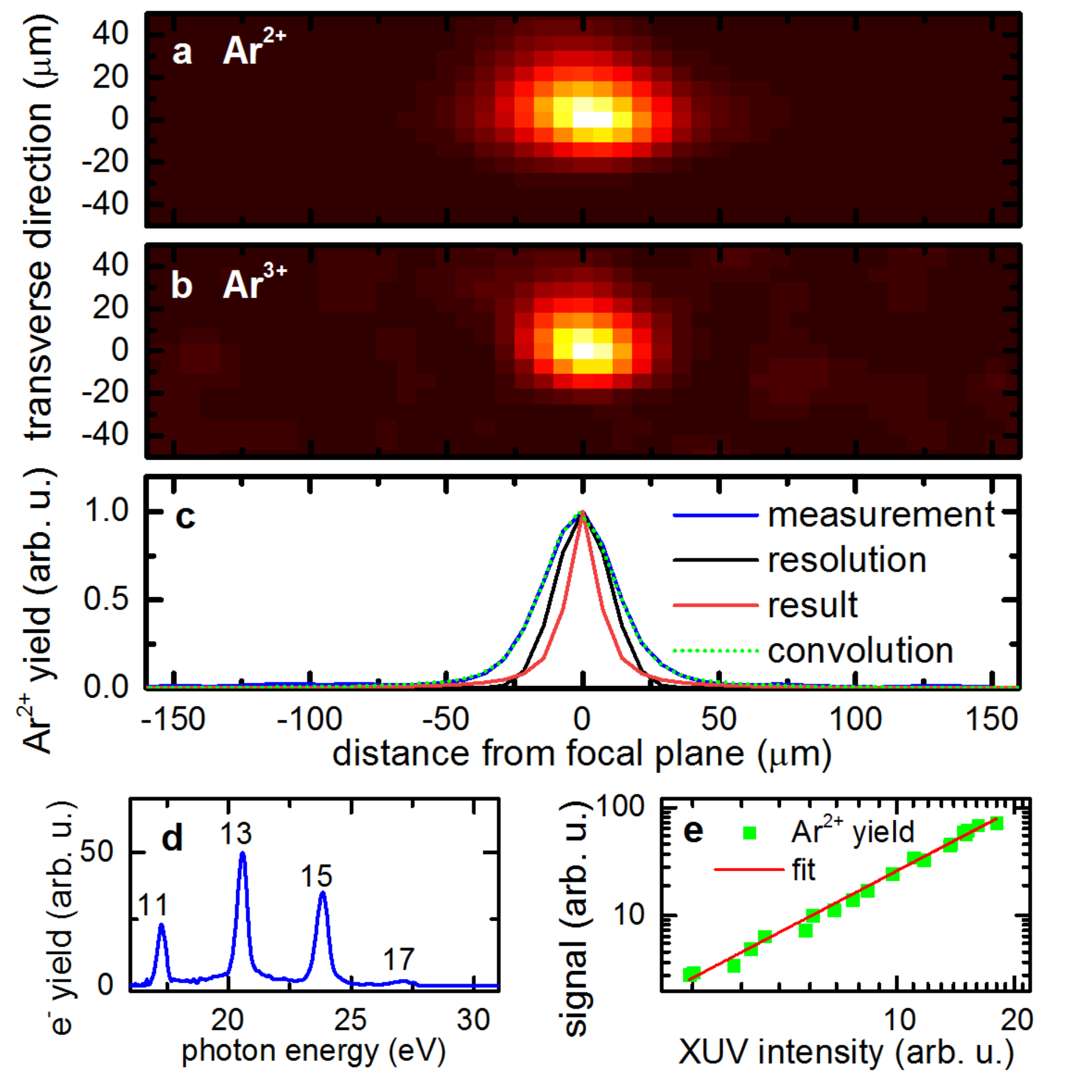}
 \caption{\label{figure_Xe2+} \textbf{XUV multiphoton ionization of Ar.} \textbf{a,} Ar$^{2+}$ and \textbf{b,} Ar$^{3+}$ ion yields as a function of the distance from the XUV focal plane. The horizontal distribution is slightly narrower for Ar$^{3+}$, reflecting the higher nonlinearity in this case. \textbf{c}, Measured Ar$^{2+}$ ion yield (blue curve), spatial resolution (black curve) and the deconvoluted Ar$^{2+}$ ion yield distribution (red curve). The convolution of the latter two contributions (green curve) agrees well with the measurement. \textbf{d}, Photoelectron spectrum following ionization of Ar atoms by the XUV pulses, showing contributions from the harmonic orders 11 to 17. \textbf{e,} Ar$^{2+}$ ion yield as a function of the XUV intensity, which is fitted by the function $A \times I^n$, resulting in $n=1.9\pm 0.1$.}
\end{figure}

The ability to generate intense XUV pulses is experimentally demonstrated by studying multiphoton absorption in Ar atoms. To this end, we generated harmonics in Xe (using a backing pressure of 2\,bar and an NIR pulse energy of 8\,mJ) and focused the XUV pulses into an atomic jet using a B$_4$C-coated spherical mirror with a focal length of 75\,mm. Applying ion spatial imaging (see Methods for details), Ar$^{2+}$ and Ar$^{3+}$ ion yields were recorded along the XUV propagation direction, as shown in Fig.~4a,b. While the Ar$^+$ ion yield is constant as a function of the distance from the XUV focal plane (not shown), the Ar$^{2+}$ and Ar$^{3+}$ ion yields are peaked at the XUV focal plane, demonstrating that these signals scale nonlinearly with the XUV intensity. The generation of Ar$^{2+}$ from neutral Ar requires an energy of at least 43.4\,eV, indicating that its generation is the result of a two-photon absorption process, considering that the maximum XUV photon energy is 27.8\,eV (see Fig.~4d). Indeed, the Ar$^{2+}$ ion yield was found to scale as $I^{1.9\pm0.1}$ (see Fig.~4e). Similarly, the generation of Ar$^{3+}$ ions, which requires a minimum energy of 84.1\,eV, is attributed to a four-photon absorption process. In combination with the measured XUV beam radius on the XUV focusing mirror ($w_{XUV,mirror}=7.4$\,mm), the Ar$^{2+}$ ion distribution along the XUV propagation direction (blue curve in Fig.~4c) can be used to estimate the Rayleigh length of the XUV beam as $z_R=6.5$\,$\mu$m (see Methods) and its waist radius as $w_0 = w_{XUV,mirror} (1 + d_{XUV}^2/z_R^2)^{-1/2} \approx w_{XUV,mirror} z_R/d_{XUV}=(600\pm 100)$\,nm, where $d_{XUV}=81$\,mm is the distance between the XUV focusing mirror and the image plane. The larger value with respect to the calculated beam waist radius of 300\,nm may be attributed to imperfect alignment as well as to wavefront distortions induced by the NIR pulse and by the XUV focusing mirror. Combining this waist radius with the afore-mentioned XUV pulse duration of 25\,fs and the pulse energy of 30\,nJ, the experimental XUV peak intensity is estimated as $I_{peak} =2\times10^{14}$\,W/cm$^2$. In the future, significantly higher XUV intensities could be achieved by further scaling of our approach. We expect the XUV intensity to scale as $I_{XUV} \propto E_{NIR}^2$ (where $E_{NIR}$ is the NIR pulse energy), if all relevant parameters including the NIR beam size before focusing and the XUV focusing mirror size are scaled appropriately. Using e.g. an NIR pulse energy of 40\,mJ instead of 8\,mJ for HHG in Xe as well as a 5 times larger NIR beam area before focusing (using the same focusing lens), the XUV pulse energy is expected to increase by a factor of 5 due to the 5 times larger generation area, whereas the NIR focal area and the XUV virtual source area are expected to decrease by a factor of 5, resulting in a 25 times higher XUV intensity. In this case, the XUV focusing mirror needs to be able to accommodate the $\sqrt{5}$ larger XUV beam divergence. If accompanied by a reduction of the XUV pulse duration $\leq 1$\,fs~\cite{tzallas11, takahashi13}, an intensity of $10^{17}$\,W/cm$^2$ would come within reach.

In summary, we have demonstrated a compact setup for the generation of high XUV intensities by generating high harmonics several Rayleigh lengths away from the driving laser focus. This concept benefits from two favorable properties: (i) A large number of XUV photons are emitted from the large generation volume without the need to apply a loose-focusing geometry. (ii) XUV pulses with curved wavefronts are generated, and are accompanied by a small virtual XUV source size, making refocusing of the XUV pulses to a small beam waist radius possible. The small size and robustness of our concept makes it straightforward to generate intense XUV pulses in a large number of laboratories in the future. Areas that we may expect to benefit from this development are XUV-pump XUV-probe spectroscopy in gases, liquids and solids as well as coherent diffractive imaging of nanoscale structures and nanoparticles. Furthermore, our concept is ideally suited for experiments that require either a high XUV flux or a small XUV focus.

\section*{References}
%\bibliography{Bibliography}


\section*{Methods}

\noindent \textbf{Experimental methods.} The experiments were performed using a Ti:sapphire laser system~\cite{gademann11} operating at a central wavelength of 790\,nm and delivering pulses with an energy up to 35\,mJ and a duration of 40\,fs. A pulse energy of up to 16\,mJ was used in the current experiments. These pulses were focused using a spherical lens with a focal length of 1\,m and coupled into the vacuum using a 3\,mm thick fused silica window. A pulsed gas jet was generated by a piezoelectric valve with a nozzle diameter of 0.5\,mm that was mounted from above, applying a backing pressure of up to 10\,bar. The relative position of the laser focus with respect to the gas jet was varied by mounting the lens on a long translation stage. A charge-coupled device (CCD) camera was used to record NIR beam profiles at the gas jet position.

The NIR pulses co-propagating with the HHG pulses were attenuated using a 100\,nm thick Al filter. The XUV pulses were spectrally resolved using a diffraction grating, and the spectra were recorded using a microchannel plate (MCP) / phosphor screen assembly in combination with a CCD camera. The XUV beam profile was measured via the same detection method and using the grating in zeroth order. The XUV pulse energy was measured by an XUV photodiode (AXUV100G) that was temporarily placed at a distance of about 0.5\,m behind the gas jet.

To generate high XUV intensities, the XUV pulses were refocused using a B$_4$C-coated spherical mirror with a focal length of 75\,mm that was placed at a distance of 103\,cm behind the gas jet. The focused XUV beam was intersected by a pulsed gas jet that was generated by a piezoelectric valve~\cite{irimia09}. A molecular beam skimmer with an orifice diameter of 0.5\,mm was used to select the central part of the atomic beam and to provide efficient differential pumping between the gas jet chamber and the interaction chamber. Photoions were generated in the interaction zone of a velocity-map imaging spectrometer~\cite{eppink97}, which was operated in spatial imaging mode. The MCP / phosphor screen detector of the VMIS was gated to be able to separately record the ions in different charge states. 

To determine the XUV Rayleigh length from the Ar$^{2+}$ ion distribution, the measurement shown in Fig.~4a was repeated at a lower gas density to avoid any possible space charge effects. The Ar$^{2+}$ ion yield scales with $I^2(z) \times w^2 (z) \propto 1/w^4(z) \times w^2(z)= 1/w^2(z)$, assuming that two XUV photons are required for the generation of Ar$^{2+}$. Here $I(z)$ is the intensity as a function of the distance from the focal plane and $w(z)=w_0\sqrt{(1+z^2/z_R^2)}$ is the beam radius as a function of the distance from the focal plane. It follows that the Ar$^{2+}$ ion distribution is proportional to $(1+z^2/z_R^2)^{-1}$. This formula was used to fit the measured Ar$^{2+}$ ion distribution. In addition, the transverse Ar$^{2+}$ ion distribution was used to determine the spatial resolution after applying a Gaussian fit. Deconvolution of these two curves results in the extracted Ar$^{2+}$ ion distribution shown as a red curve in Fig.~4c, from which a Rayleigh length of 6.5\,$\mu$m was obtained. For comparison, we also show the convolution of the red and black curves as green dotted curve, which agrees well with the measured Ar$^{2+}$ distribution (blue curve).

\vspace{5mm}

\noindent \textbf{Numerical methods.} 
The HHG simulations were performed with an extended version of a three-dimensional nonadiabatic simulation code described in~\cite{tosa03}. The model solves the paraxial wave equation in combination with the Lewenstein integral~\cite{lewenstein94} to obtain the macroscopic response of a gas medium to a strong laser field. As a first step, propagation of the laser pulse in the ionized medium is calculated assuming a cylindrical symmetry and taking into account the time- and space-dependent neutral and plasma dispersion, along with the optical Kerr effect. The dipole responses of single atoms in the interaction volume are obtained using the strong-field approximation. These dipole responses serve as source terms in each spatial grid point of the interaction medium when solving the wave equation of similar form as for the fundamental laser field. When calculating the propagation of the harmonic field, neutral dispersion and absorption are taken into account. Further details are available in the supplementary material of Ref.~\cite{rivas18}.
Backpropagation of the generated high-harmonic radiation was carried out using the ABCD-Hankel transform~\cite{major18}.

\vspace{5mm}

%\section*{Data Availability Statement}

%The data that support the findings of this study are available from the corresponding author upon reasonable request.

\section*{Acknowledgements}

We thank M. Krause, R. Schumann and C. Reiter for their support with the laser system. The ELI-ALPS project (GINOP-2.3.6-15-2015-00001) is supported by the European Union and co-financed by the European Regional Development Fund. We acknowledge KIF\"U for awarding us access to HPC resources based in Hungary. K. Kov\'acs and V. Tosa acknowledge support from the grant ELI\_03 Pulse-MeReAd.

\section*{Author contributions}

B.M. and O.G. contributed equally to this work. B.S. and O.G. performed the experiments. B.M., K.K. and V.T. carried out the calculations. All authors discussed the results and contributed to writing the manuscript.

\section*{Competing financial interests}

The authors declare no competing financial interests.

\section*{Additional information}

%\noindent \textbf{Supplementary Information} is available for this paper at ...

\noindent \textbf{Correspondence and requests for materials} should be addressed to B.S. or B.M.

\end{document}